\documentclass[%
 reprint,
 superscriptaddress,
 amsmath,amssymb,
 aps
]{revtex4-2}

\usepackage{graphicx}% Include figure files
\usepackage{dcolumn}% Align table columns on decimal point
\usepackage{bm}% bold math
\usepackage{braket}
\usepackage{qcircuit}
\usepackage{hyperref}% add hypertext capabilities

\newcommand{\iu}{i}
\DeclareMathOperator{\Tr}{Tr}

\begin{document}

\title{Quantum supremacy regime for compressed fermionic models}

\author{Guillermo Bl\'{a}zquez-Cruz}
\email{guilleblazquez@gmail.com}
\affiliation{
Quantum Open Source Foundation, Toronto, Canada
}
\author{Pierre-Luc Dallaire-Demers}
\affiliation{
Quantum Open Source Foundation, Toronto, Canada
}
\affiliation{
Pauli Group Inc., Toronto, Canada
}

\date{\today}

\begin{abstract}
Compressible models extend the domain of simulable systems in quantum computers, but little is known about their precise limits of applicability. Using the theory of compressible matchgate circuits, we identify a class of quadratic fermionic Hamiltonians that can be simulated in compressed space. In particular, for systems of $n$ orbitals encoded to 2-local qubit models with nearest neighbour interactions, the ground state energy can be evaluated with $O\left(\log n\right)$ sets of measurements, independently of the number of dimensions in which the $n$ sites are arranged. We also provide an expressible circuit ansatz in a logarithmic number of qubits for finding the compressed ground state with a variational quantum eigensolver. From the complexity analysis of the compressed circuits, we find a regime of quantum supremacy for sampling compressed Gaussian fermionic models.
\end{abstract}

\maketitle

\section{\label{sec:intro}Introduction}
The simulation of quantum physical systems has been one of the most anticipated applications of quantum computation since Feynman's seminal paper \cite{Feynman1982}. Current devices, so-called noisy intermediate-scale quantum (NISQ) computers, suffer from practical limitations that restrict the number of qubits and their coherence. Despite these limitations, several algorithms have been proposed that overcome these resource constraints. Amongst these, the variational quantum eigensolver (VQE) \cite{Peruzzo2014} has found applications in the simulation of many-body physics, quantum chemistry or combinatorial problems (see \cite{Bharti2021} for a review of NISQ algorithms and their applications). In a VQE algorithm, a parameterized circuit $U\left(\boldsymbol{\theta}\right)$ is minimized with respect to an objective function, usually the expectation value of the energy of a Hamiltonian $H$,
\begin{equation}
\min_{\boldsymbol{\theta}} \Braket{0|U^\dagger\left(\boldsymbol{\theta}\right) H U\left(\boldsymbol{\theta}\right)|0},
\end{equation}
with the expectation value computed in a quantum device and the minimization procedure performed classically.

A particularly interesting class of many-body quantum systems are those that can be mapped to non-interacting fermions, i.e., their Hamiltonians are quadratic in the fermionic operators. Their eigenstates in the fermionic picture are single-mode non-interacting states and, in many cases, the modes can be found analytically. Despite their simplicity, quadratic Hamiltonians appear in a variety of settings. They arise as mean field approximations to higher order Hamiltonians such as the BCS model of superconductivity \cite{Bardeen1957} or the Fermi-Hubbard model \cite{1963a}. They also contain the simplest examples of topological insulators like the Kitaev wire \cite{Kitaev2001} or the SSH model \cite{Su1979, Su1980}. In some algorithms, they are used as a starting point to simulate interacting fermions, for example in the Fermi-Hubbard model \cite{Wecker2015, Montanaro2020} or in Hartree-Fock settings \cite{DallaireDemers2018}. Other simpler many-body models with strong correlations like the XY chain or the 1D Ising model are also described by quadratic Hamiltonians and have been solved analytically or in quantum computers \cite{Verstraete2008, CerveraLierta2018}.

Quadratic interactions are equivalent to matchgate circuits, and can be simulated in logarithmic space. The notion of compressed quantum computation has been used in NISQ devices to overcome the memory limitation of current computers. Matchgate based compression has been used to simulate the 1D Ising and XY models numerically \cite{Kraus2011, Boyajian2013, Boyajian2015}, and experimentally \cite{Hebenstreit2017, CerveraLierta2018, Li2014}.

In this paper we study the compressed simulation of free fermionic systems from the perspective of variational state preparation algorithms. In section \ref{sec:compressed} we review the main ideas of matchgate circuits and their compression. In section \ref{sec:ansatz} we define a class of parametrized quantum circuits (PQC) ansatz with up to a quadratic number of parameters in the system size. In section \ref{sec:operators} we show that the expectation value of a compressed quadratic operator with nearest-neighbor interactions can be computed in a logarithmic number of sets of measurements. The complexity analysis of the full circuits reveals a regime of quantum supremacy for compressed fermionic models. Finally, in section \ref{sec:conclusion} we discuss the results.

\section{\label{sec:compressed}Fermionic gaussian states and their compression}
Let us first recall the main ideas in the compression of fermionic Gaussian states, to illustrate how to run a VQE algorithm in compressed space. We refer the reader to Jozsa et al. \cite{Jozsa2008, Jozsa2009} and Boyajian \cite{Boyajian2015a} for details.

A quadratic fermionic Hamiltonian on $n$ orbitals can be written as
\begin{equation}
    H = \iu \sum_{j \neq k = 0}^{2n - 1} h_{j,k} \gamma_j \gamma_k,
    \label{eq:quadratic_operator}
\end{equation}
where the $\gamma$'s are Majorana operators and $h$ is a real, antisymmetric matrix. The unitaries generated by these Hamiltonians are called Gaussian. Majorana operators transform under conjugation by Gaussian operators $U$ as
\begin{equation}
U^\dagger \gamma_j U = \sum_{k = 0}^{2n-1}R_{j,k}\gamma_k,    
\label{eq:MajoranaOperatorsConjugationGaussian}
\end{equation}
with $R = e^{4h} \in SO\left(2n\right)$. A Gaussian transformation on $n$ orbitals is expressible as a circuit of $O\left(n^2\right)$ nearest-neighbour matchgates on $n$ qubits. A matchgate is a quantum gate of the form
\begin{equation}
    \mathcal{G}\left(A, B\right) = \begin{pmatrix}
    p & 0 & 0 & q \\
    0 & w & x & 0 \\
    0 & y & z & 0 \\
    r & 0 & 0 & s
    \end{pmatrix},
    u = \begin{pmatrix}
    p & q \\ r & s
    \end{pmatrix},
    v = \begin{pmatrix}
    w & x \\ y & z
    \end{pmatrix}
\end{equation}
with $u$ and $v$ unitary matrices satisfying $\det\left(u\right) = \det\left(v\right)$.
A nearest-neighbour matchgate circuit on $n$ qubits can either be simulated efficiently in a classical computer \cite{Valiant2002, Jozsa2008}, or in a quantum computer with $\log{n} + 1$ qubits \cite{Jozsa2009}, provided that we measure a quadratic operator.

The ground states of quadratic fermionic Hamiltonians are called fermionic Gaussian states. They can be written as $\omega = K e^{-\frac{\iu}{4} \mathbf{x}^T G \mathbf{x}}$, where $\mathbf{x}$ is a vector of Majorana operators, $G$ is a real, antisymmetric matrix and $K$ is a normalization factor. We are interested in the compression of this set of states and the Gaussian operators. A fermionic Gaussian state $\omega$ is completely characterized by its second moments, encoded in the covariance matrix
\begin{equation}
    \Gamma\left(\omega\right)_{kl} = \frac{\iu}{2}\Tr\left(\omega {[\gamma_k, \gamma_l]} \right ).
    \label{eq:covariance_matrix}
\end{equation}
All information about the state can be obtained from $\Gamma$ using Wick's theorem. The covariance matrix can be used to define a density matrix $\rho$ on $\log{n} + 1$ qubits as
\begin{equation}
    \rho\left(\omega\right) = \frac{1}{2n}\left( \mathbb{I} + \iu\Gamma\left(\omega \right ) \right ).
    \label{eq:compressed_state}
\end{equation}
Fermionic Gaussian states are closed under the action of Gaussian operators $U$. By Eq. \ref{eq:MajoranaOperatorsConjugationGaussian}, the corresponding action on $G$, $\Gamma$ and $\rho$ is the conjugation by the rotation matrix $R$, which takes the form of a Bogoliubov transformation. Finally, the expectation value of an operator $H$ can be computed in compressed space as
\begin{equation}
    \braket{H} = n \Tr\left[ R \rho R^T A \right ] 
\label{eq:compressed_expectation_value}
\end{equation}
where $A = ih$.

We have now all the necessary steps to run a VQE algorithm in compressed space: start with any compressed fermionic Gaussian state, parametrize an $SO\left(2n\right)$ rotation to explore the whole set of compressed fermionic Gaussian states and measure the compressed operator $A$ to get the expectation value $\braket{H}$.

Ideally, we would like to grow the uncompressed number of orbitals exponentially and simulate them in $O\left(\log{n}\right)$ qubits and sets of measurements. In the rest of the paper we explain how to implement an expressible, although quadratic, PQC ansatz and we identify a class of Hamiltonians for which the compressed operator can be efficiently computed.

\section{\label{sec:ansatz} Parametrized quantum circuits for compressed Fermionic Gaussian states}
In this section, we explain how a PQC ansatz for compressed fermionic models can be built to parametrize Givens rotations both individually and in large blocks.

Quadratic Hamiltonians have fermionic Gaussian states as ground states. These states are closed under the action of Bogoliubov transformations, or, in compressed space, special orthogonal rotations. Same parity states are closed under the action of proper Bogoliubov transformations (having determinant 1), while different parity sectors are connected by an improper transformation, with determinant -1. Hence, a PQC ansatz that implements a special orthogonal rotation will be able to explore the set of even or odd compressed Fermionic Gaussian states, if the initial state is itself a member of that set. From now on, we use $n$ to denote the number of orbitals in the system and $m = \log{n} + 1$ for the number of qubits in the compressed simulation.

Here we explain how to generate the compressed version of the $\ket{0}^{\otimes n}$ state. Its covariance matrix is defined by
\begin{equation}
\Gamma\left(\ket{0}^{\otimes n}\right)_{j,k} = \begin{cases}
-1 & \text{ if } (j,k) = (2l, 2l + 1)\\ 
1 & \text{ if } (j,k) = (2l + 1, 2l)\\ 
0 & \text{ otherwise }
\end{cases}
    \label{eq:zero_state_covariance_matrix}
\end{equation}
with $l \in \mathbb{Z}$, $0 \leq l \leq n - 1$. The associated $m$-qubit density matrix is
\begin{equation}
\rho = \frac{1}{n}\left( \mathbb{I}  \otimes \ket{+_y}\bra{+_y} \right )
    \label{eq:zero_compressed_density_matrix}
\end{equation}
In general, compressed fermionic Gaussian states are mixed. We can purify the state in Eq. \ref{eq:zero_compressed_density_matrix} in $2m - 1$ qubits as
\begin{equation}
    \ket{\psi}= \frac{1}{\sqrt{2n}}\sum_{j=0}^{2^{m - 1} - 1}{\ket{j}_\mathcal{L} \ket{0} \ket{j}_\mathcal{R} +\iu\ket{j}_\mathcal{L} \ket{1} \ket{j}_\mathcal{R}}.
    \label{eq:zero_compressed_state}
\end{equation}
Tracing out subsystem $\mathcal{R}$ generates the density matrix in Eq. \ref{eq:zero_compressed_density_matrix}. The previous state can be prepared in a quantum computer with the following $2m - 1$ qubit circuit,
\[
\Qcircuit @C=.7em @R=1.0em {
\lstick{\ket{0}} & \qw & \qw & \qw & \qw & \targ & \qw \\
& \vdots & & & & & \\
\lstick{\ket{0}} & \qw & \qw & \targ & \qw & \qw & \qw\\
\lstick{\ket{0}} & \gate{H} & \gate{S} & \qw & \qw & \qw & \qw\\
\lstick{\ket{0}} & \gate{H} & \qw & \qw & \qw & \ctrl{-4} & \qw\\
& \vdots & & & & & \\
\lstick{\ket{0}} & \gate{H} & \qw & \ctrl{-4} & \qw & \qw & \qw
}
\]
where the $S$ gate is in the $m$-th qubit.

With a suitable initial state like this one, we can rotate the relevant degrees of freedom using a special orthogonal rotation. Any $SO(2n)$ element can be decomposed in $n(2n-1)$ Givens rotations, which are rotations around a single axis and take the form
\begin{equation}
    R_{i,j}\left(\theta \right )=\begin{pmatrix}
    1 & \ldots & 0 & \ldots & 0 & \ldots & 0 \\ 
    \vdots & \ddots & \vdots &  & \vdots &  & \vdots \\ 
    0 & \ldots & \cos \theta& \ldots & -\sin \theta & \ldots & 0 \\ 
    \vdots &  & \vdots & \ddots  & \vdots &  & \vdots \\ 
    0 & \ldots & \sin \theta & \ldots & \cos \theta & \ldots & 0 \\ 
    \vdots &  & \vdots &  & \vdots & \ddots  & \vdots \\ 
    0 & \ldots & 0 & \ldots & 0 & \ldots & 1
    \end{pmatrix},
    \label{eq:givens_rotation}
\end{equation}
where the trigonometric functions appear at the intersections of the $i$-th and $j$-th rows and columns. To completely parametrize an $SO(2n)$ element, we can restrict to Givens rotations with $i < j$. Consider first the case in which the binary strings of $i$ and $j$ differ by just one position with index $k$. A multicontrolled RY gate on qubit $k$ controlled by all other other qubits being on implements a Givens rotation between states $\ket{1\ldots 101\ldots 1}$ and $\ket{1\ldots 111\ldots 1}$, where the $0$ and the respective $1$ are at position $k$. To implement a more general rotation, all qubits at positions other than $k$ must be turned into a 1 before the RY gate and then back to a 0, which can be done in at most 4 gates per qubit. A fully-controlled RY gate in $m$ qubits can be implemented in $O\left(m\right)$ elementary operations \cite{Barenco1995}. The total complexity of the PQC ansatz is $O\left(n^2\log(n)\right)$, with the quadratic dependence coming from the number of Givens rotations. We provide a Python package that builds this PQC ansatz for Qiskit \cite{BlazquezCruz2021} \footnote{Note that although the ansatz is the one we propose, the measurement procedure in Qiskit's VQE algorithms uses the Pauli words instead of the procedure explained in the next sections. The package can also convert a quadratic 1D Hamiltonian into its compressed Pauli words as an example.}.

Odd states can be explored turning an even state into an odd one with a determinant $-1$ unitary, such as the fully-controlled $X$, with no effect in the PQC complexity.

This PQC ansatz has $n\left(2n - 1\right)$ parameters and can explore all the set of $SO\left(2n\right)$ rotations. In practice, the ansatz is too expressive, the ground state can be found by exploring rotations affecting only specific elements in the correlation matrix. We postulate that there exists polynomial size PQC (therefore polynomial quantum runtime) that generates valid Givens rotations over the exponentially large uncompressed space. For example, an RY rotation on qubit $k$ out of $m$ decomposes into Givens rotations as
\begin{equation}
RY(\theta)=\prod_{i=0}^{2^k-1}\prod_{j=0}^{2^{m-k-1}-1}R_{i2^{m-k}+j,i2^{m-k}+j+2^{n-k-1}}(\theta)    
\end{equation}
Therefore, a PQC with one RY per qubit would generate a polynomial size ansatz encoding $O(n\log(n))$ Givens rotations, whose space of solutions is constrained by the polynomial number of parameters. It is an open question which parent Hamiltonians can these polynomial size PQC solve.

Even assuming polynomial PQCs for interesting Hamiltonians, computing expectation values in compressed space on a quantum computer is not immediate. In the next section we reduce this problem for quadratic operators $\gamma_j\gamma_k$ to that of sampling from a state defined by the PQC and an operator depending on $j$ and $k$, and show that the number of such operators is proportional to the number of qubits $O(m)$.

\section{\label{sec:operators}Efficiently Measurable Compressed Operators}

The cost of measuring an arbitrary compressed Hamiltonian when the number of orbitals $n$ grows exponentially can be extremely high, because a compressed operator has up to $n(2n - 1)$ terms in its Pauli basis expansion. In this section we first show how to group the set of Pauli words of any compressed operator in sets of mutually commuting operators. Then, we show that nearest-neighbour interactions reduce the number of sets exponentially, even if we stack the fermions in any number of dimensions. We give the diagonalizing circuits. Finally, we analyze the classical complexity of sampling the same observables and identify a regime of quantum supremacy.

The compressed operator $A$ defined in Section \ref{sec:compressed} is structured in $n^2$ blocks $a_{j,k}$ of dimensions $2\times 2$,
\begin{equation}
a_{j,k} = i\begin{pmatrix}
h_{2j,2k} & h_{2j, 2k+1} \\
h_{2j+ 1, 2k} & h_{2j+ 1, 2k + 1}
\end{pmatrix}.
\end{equation}
There is a mapping between the submatrices $a_{j,k}$ and operations between orbitals in the Pauli basis. To make it evident, we choose the Jordan-Wigner representation for Majorana operators on $n$ qubits in terms of Pauli matrices, $\gamma_{2j} = \left(\bigotimes_{k=0}^{j-1} \sigma_z\right) \sigma_x$ and $\gamma_{2j + 1} = \left(\bigotimes_{k=0}^{j-1} \sigma_z\right) \sigma_y$ for $j \in \lbrace 0, \ldots, n-1 \rbrace$. With this representation, a quadratic operator is
\begin{equation}
\gamma_k \gamma_l = \begin{cases}\iu\sigma_z^\alpha, &\text{$k=2\alpha$, $l = 2\alpha+1$} \\
-\iu\sigma_y^\alpha\sigma_z^{\alpha+1}\ldots\sigma_z^{\beta-1}\sigma_x^\beta, &\text{$k=2\alpha$, $l = 2\beta$} \\
-\iu\sigma_y^\alpha\sigma_z^{\alpha+1}\ldots\sigma_z^{\beta}\sigma_y^\beta, &\text{$k=2\alpha$, $l = 2\beta + 1$} \\
\iu\sigma_x^\alpha\sigma_z^{\alpha+1}\ldots\sigma_z^{\beta-1}\sigma_x^\beta, &\text{$k=2\alpha + 1$, $l = 2\beta$} \\
\iu\sigma_x^\alpha\sigma_z^{\alpha+1}\ldots\sigma_z^{\beta-1}\sigma_y^\beta, &\text{$k=2\alpha + 1$, $l = 2\beta + 1$} \\
\end{cases}
\end{equation}
with $k < l$ and where the superscript labels the site. One can see that the diagonal blocks $a_{k,k}$ encode $n$ single qubit $\sigma_z$ operators. The non-diagonal blocks $a_{k,l}$ encode the four possible interactions between orbitals $k$ and $l$.

\subsection{Sets of mutually commuting operators of an antisymmetric operator}

An antisymmetric $2n \times 2n$ matrix has $n(2n-1)$ free parameters, and its Pauli basis expansion will have at most the same number of words. Only antisymmetric tensor products (i.e., with an odd number of $\sigma_y$'s), can be present in the expansion of an antisymmetric matrix. Let $\sigma_A = \lbrace \sigma_x, \sigma_y\rbrace$ represent the set of antidiagonal Pauli matrices, and $\sigma_D = \lbrace \mathbb{I}, \sigma_z\rbrace$. Let $\sigma_{M}$ be any Pauli matrix $\sigma_M \in \lbrace \sigma_D, \sigma_A \rbrace$. Then, the tensor product $\bigotimes_{i=1}^m \sigma_{M_i}$ with at least one $\sigma_{M_i} \in \sigma_A$ is a set of mutually commuting observables. There are $2n-1$ such tensor products, and $n$ elements in each product, for a total of $n(2n-1)$ parameters. We show in the Supplemental Material \ref{sec:Scommuting} that each of these products form a set of mutually commuting observables. Fig. \ref{fig:8orbitals} (a) exemplifies the relationship between sets of tensor products, $a_{i,j}$ blocks and orbital interactions for an 8-orbital system.

With these sets we can compute any compressed operator's expectation value in $2n - 1$ sets of measurements. To find classes of Hamiltonians where a logarithmic number of sets of measurements is needed we must relate each tensor product to the matrix elements it can populate. The non-zero elements of (omitting the tensor product symbol), $\sigma_{M_1} \sigma_{M_2} \ldots \sigma_{M_m}$, can be tracked by starting at the center of the matrix and selecting diagonal or antidiagonal subblocks according to the label $M_i$. For example, for a $4 \times 4$ matrix, the $\sigma_A\sigma_D$ set will have the non-zero elements
\begin{eqnarray*}
    \left(
    \begin{array}{c|c}
        \begin{array}{cc}
            h_{0,0} & h_{0,1} \\
            h_{1,0} & h_{1,1}
        \end{array} & \begin{array}{cc}
            h_{0,2} & h_{0,3} \\
            h_{1,2} & h_{1,3}
        \end{array} \\ \hline
        \begin{array}{cc}
            h_{2,0} & h_{2,1} \\
            h_{3,0} & h_{3,1}
        \end{array} & \begin{array}{cc}
            h_{2,2} & h_{2,3} \\
            h_{3,2} & h_{3,3}
        \end{array}\end{array}
    \right) \\ \xrightarrow{A}
        \left(
    \begin{array}{cc}
         & \begin{array}{c|c}
            h_{0,2} & h_{0,3} \\ \hline
            h_{1,2} & h_{1,3}
        \end{array} \\ 
        \begin{array}{c|c}
            h_{2,0} & h_{2,1} \\ \hline
            h_{3,0} & h_{3,1}
        \end{array} & \end{array}
    \right)  \\ \xrightarrow{D}
        \left(
    \begin{array}{cc}
         & \begin{array}{cc}
            h_{0,2} &  \\ 
             & h_{1,3}
        \end{array} \\ 
        \begin{array}{cc}
            h_{2,0} &  \\
             & h_{3,1}
        \end{array} & \end{array}
    \right) 
\end{eqnarray*}
Similarly, we can trace back the set of tensor products from a matrix element by following the path backwards.

\subsection{Compressed operators with nearest-neighbour interactions in arbitrary dimensions}
While the grouping of Pauli words in sets of commuting operators reduces the number of measurements, there is still an exponential growth with the number of compressed qubits. We now consider the Pauli basis expansion of operators with nearest-neighbour (n.n.) interactions, including many well-known systems such as the tight-binding models or the Kitaev wire. The structure of the compressed operator is such that the number of sets of measurements to be made in a quantum computer is reduced exponentially.
\begin{figure}
\includegraphics{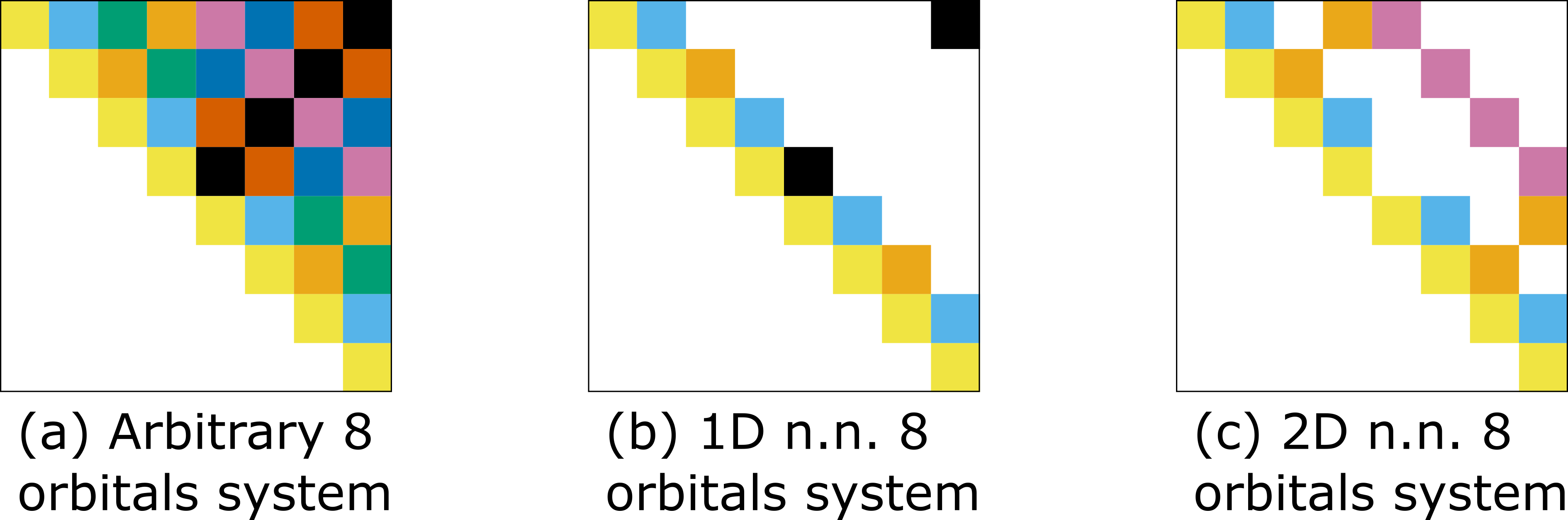}
\caption{\label{fig:8orbitals} Schematic structure of a quadratic Hamiltonian compressed operator for a system of 8 orbitals. Each colored square at position $i,j$ is a $2$x$2$ submatrix encoding interactions between orbitals $i$ and $j$. The different shades of colors are used to identify the sets of elements that are spanned by a given common set of mutually commuting Pauli words. E.g., black squares encode interactions between orbitals $0-7$, $1-6$, $2-5$ and $3-4$, and are spanned by the sets $\sigma_A\sigma_A\sigma_A\sigma_M$, where $M \in \lbrace D, A\rbrace$ span the diagonal or antidiagonal black-square elements, respectively. We only color the upper triangular elements because of the operator's antisymmetry. White blocks in the upper triangular section are $0$. (a) all possible interactions of an 8-orbital system. The number of sets is linear in the orbitals. (b) interactions of a 1D nearest-neighbor system. The number of sets is logarithmic in the orbitals. The $0-7$ interaction is only included with periodic boundary conditions. (c) interactions of a 2D n.n. system. Black squares have been replaced by pink squares. The new $0-3$ and $4-7$ interactions are periodic boundary conditions. The number of sets is still logarithmic in the number of orbitals.}
\end{figure}

Let us start with a 1D system of size $n$ and treat open boundary conditions first. In this case, only the $a_{i,i}$ and $a_{i, i+1}$ will be non-zero. The $a_{i,i}$ terms are spanned by $\sigma_D\ldots\sigma_D\sigma_A$ and encode local $\sigma_z$'s. The $a_{i,i+1}$ encode n.n. interactions. We can collect them in $2\left(m - 1\right)$ sets of commuting observables. Let $l = 1,\ldots, m-1$ and, for each $l$, $k = 0, \ldots, n/2^l - 1$. The $a_{i,i+1}$ blocks with $i = 2^l k + 2^{l-1} -1$ are spanned by the sets of words (using subscripts to label the qubits) $\sigma_{D_1}\ldots\sigma_{D_{ m - l - 1}}\sigma_{A_{m - l}}\ldots\sigma_{A_{m-1}}\sigma_M$. The last label differentiates between the two possible sets of interactions in any $a_{i,j}$. Since $l$ completely labels the sets of tensor products, there are $2\left(m-1\right)$ sets in total. Periodic boundary conditions include the $a_{0, n-1}$ block, which is spanned by the same set than $a_{n/2 - 1, n/2}$. See Fig. \ref{fig:8orbitals} (b) for an 8-orbital example.

Let us turn the system two-dimensional. We can start with the 1D system and break it in equal parts at positions $n/2 - 1$ and $n/2$. The compressed matrix when there is no interaction between both halves of a chain is block diagonal, with two $n/2 \times n/2$ blocks. Compared to the connected 1D chain, the Pauli basis expansion lacks the $\sigma_A\ldots\sigma_A\sigma_M$ set. When connecting both halves to make a 2D system, this set will now be replaced by the interactions between positions $i$ and $n/2 + i$ for $i=0,\ldots, n/2 - 1$. All these interactions are spanned by the set $\sigma_A\sigma_D\ldots\sigma_D\sigma_M$. Therefore, folding a chain by the middle position does not increase the number of sets of measurements. We provide an example in Fig. \ref{fig:8orbitals} (c).

In general, if we want to fold a 1D chain into a 2D system of of $p$ rows and $q$ columns, we can start by breaking the chain in half, eliminating one set of Pauli words, then breaking the resulting two chains in half again, and keep going until we have $p$ chains of length $n/p$, at which point we have eliminated $\log{p}$ sets. Then, we can fold the smallest chains in pair by the second dimension, introducing a single set at each step, until we have introduced $\log{p}$ sets again. The same reasoning can be made to fold a chain in more than 2 dimensions.

This shows that any nearest-neighbour quadratic Hamiltonian can be measured in compressed space in $O\left(\log n \right)$ sets of measurements. Still, there is an exponential number of words in each of these sets. In the next section we show how to compute the expectation values without knowledge of each Pauli words' coefficients.

\subsection{Tractable expectation values of quadratic operators}
Wick's theorem relates any expectation value of products of Majorana operators to elements of the covariance matrix. For quadratic operators, the expectation value is
\begin{equation}
\Braket{\gamma_k \gamma_l} = -\iu \Gamma_{k,l} = -n\Tr\left[\rho\left(\ket{l}\bra{k} - \ket{k}\bra{l}\right)\right],
\end{equation}
where $\rho$ is the compressed state. The $\ket{l}\bra{k} - \ket{k}\bra{l}$ operator is a linear combination of the Pauli words in the set of mutually commuting operators which are non-zero at position $k,l$. With the appropriate diagonalization, we can compute $\Braket{\gamma_k \gamma_l}$ directly from measurement results.

The following circuit diagonalizes the $\sigma_{A} \sigma_{A} \ldots \sigma_{A}$ set (proof in the Supplemental Material \ref{sec:Sdiagonalization}):
\[
\Qcircuit @C=.7em @R=1.0em {
& \ctrl{3} & \ctrl{1} & \gate{X} & \gate{S} & \gate{H} & \ctrl{1} & \ctrl{3}  & \qw \\
& \qw & \targ & \qw & \qw & \qw  & \targ & \qw & \qw  \\
&&&& \vdots &&&& \\
& \targ & \qw & \qw & \qw & \qw & \qw & \targ & \qw
}\]
More general tensor products are diagonalized with this circuit applied to all $\sigma_A$ positions. In terms of the matrix elements, the operator moves element $k,l$ to position $k,k$ with an $\iu$ factor. Let $V$ be the diagonalizing circuit operator, then
\begin{eqnarray}
\Gamma_{k,l} &&= -\iu n \Tr\left(V\rho V^\dagger V \left(\ket{l}\bra{k} - \ket{k}\bra{l}\right) V^\dagger\right) \nonumber \\ &&= n\Tr\left(V\rho V^\dagger \left(\ket{l}\bra{l} - \ket{k}\bra{k}\right)\right) \nonumber \\ &&= n\left(P\left(l\right) - P\left(k\right)\right),
\end{eqnarray}
where $P\left(k\right)$ and $P\left(l\right)$ are the probabilities of measuring bitstrings $k$ and $l$ in the $V\rho V^\dagger$ density matrix. With the diagonalizing circuits, we can compute expectation values of quadratic operators by sampling from the quantum computer, without needing to explicitly compute the Pauli expansion of the compressed operator.

\subsection{Analysis of the procedure complexity}
Several comments can be made regarding the complexity of the compressed VQE procedure and its classical counterpart. 
First, the sampling complexity as a function of the accuracy $\epsilon$ is the same both in the quantum case and in the classical case. As is typical in general VQE settings \cite{Gonthier2020}, in compressed VQE setting the number of measurements scales as $O\left(1/\epsilon^2\right)$. This is the same in the uncompressed case \cite{Valiant2002, Jozsa2008} whether the sampling is done classically or on a quantum device. It is also the case when the angles of the PQC are such that the ansatz can be compiled to Clifford gates and the whole circuit is Clifford \cite{Aaronson2004}.

Even in this last case where the whole circuit is Clifford, there is at least a polynomial separation between the compressed circuit runtime and its efficient classical simulation. Indeed, simulating the measurements and state updates of a Clifford circuit on $n$ qubits on a stabilizer state input requires between $O\left(n^2\right)$ and $O\left(n^3\right)$ time \cite{Aaronson2004}. However, for typical compressed VQE instances, the diagonalizing circuits are Clifford but the input states of the variational procedure defined in Sec. \ref{sec:ansatz} are not stabilizer states. Hence, the compressed quantum circuits cannot be simulated using the Gottesman-Knill procedure \cite{Aaronson2004} which means that the quantum advantage is at least quadratic in time.

If we "uncompress" the representation to a large matchgate circuit then an efficient simulation exists \cite{Valiant2002, Jozsa2008} as samples from the probability distribution can be obtained in a time polynomial in the register size. However, this uncompressed register is exponentially larger than the original quantum circuit and the total simulation time also has to be exponential. The equivalent classical computation of an expectation value involves the calculation of the Pfaffian of a matrix whose size is exponential in the number of compressed qubits. Even if the Pfaffian of a matrix can be computed in a time polynomial in the number of entries, the exponential size of the matrix implies an exponential runtime for the classical simulation.

This hints at the possibility of a quantum supremacy regime for Gaussian fermion sampling in the compressed regime. This would be an interesting complement to the supremacy regime of boson sampling \cite{Aaronson2010,Hamilton2016,Arute2019}.

Finally, there is a caveat regarding the precise classical complexity, as choosing a polynomial size cover for the domain of the PQC ansatz could leave room for a subexponential representation of the corresponding uncompressed matchgate circuit. It would imply the existence of a procedure to compute the Pfaffian of an exponentially large matrix in subexponential time. 

\section{\label{sec:conclusion}Conclusion}
We have shown how to define VQEs in compressed space and quadratic ans\"{a}tze to find ground states of quadratic Hamiltonians. The exponential space compression is accompanied by an equivalent expansion in the time complexity due to the mapping between quadratic operators and the Pauli words of the new compressed equivalents. We have shown that this can be alleviated for many physically interesting Hamiltonians thanks to two strategies. First, for any nearest-neighbour Hamiltonian, the number of sets to be measured is reduced exponentially. The interactions between fermions can include non-local terms in a way that allows any rectangular grid arrangement and open and closed boundary conditions. Second, we have shown how to compute expectation values directly from measurement results. We analyzed the complexity of the procedure and found a plausible regime of quantum supremacy for compressed free fermionic systems.

The compressed VQE could be used to simulate large-scale free-fermionic systems in NISQ devices, opening the possibility of appending compressed systems as fermionic environmental degrees-of-freedom using few qubits. Another possible use is as a first step for algorithms initialized with compressed ground states. There are algorithms that start with compressed states based on other schemes (\cite{Montanaro2020}) and algorithms starting with non-compressed fermionic Gaussian states, for example for strongly correlated systems \cite{Jiang2018} or Hartree-Fock methods \cite{DallaireDemers2018}. Other interacting systems might be studied with non-Gaussian states as input to Gaussian circuits \cite{Hebenstreit2019}. The study of compressed fermionic Gaussian states can help developing other compressed algorithms.

\begin{acknowledgments}
We wish to acknowledge the support of the Quantum Open Source Foundation in organizing the Quantum Computing Mentorship Program.
\end{acknowledgments}

\bibliography{apssamp}

\widetext
\newpage
\clearpage
\begin{center}
\textbf{\large Supplemental Material for "Quantum supremacy regime for compressed fermionic models"}
\end{center}
%%%%%%%%%% Merge with supplemental materials %%%%%%%%%%
%%%%%%%%%% Prefix a "S" to all equations, figures, tables and reset the counter %%%%%%%%%%
\setcounter{equation}{0}
\setcounter{figure}{0}
\setcounter{table}{0}
\setcounter{page}{1}\setcounter{section}{0}
\renewcommand{\theequation}{S\arabic{equation}}
\renewcommand{\thefigure}{S\arabic{figure}}
\renewcommand{\bibnumfmt}[1]{[S#1]}
\renewcommand{\citenumfont}[1]{S#1}
\renewcommand{\thesection}{S-\Roman{section}}
%%%%%%%%%% Prefix a "S" to all equations, figures, tables and reset the counter %%%%%%%%%%

\section{\label{sec:Scommuting}Construction of sets of mutually commuting tensor products for an antisymmetric operator}
Here we show that the sets of tensor products defined in Section \ref{sec:operators} are sets of mutually commuting observables. Let $P$ and $Q$ be two operators built from $m$ tensored Pauli matrices, $P = \bigotimes_{i=0}^{m-1} p_i$ and $Q = \bigotimes_{i=0}^{m-1} q_i$, with $p_i, q_i \in \sigma_D \cup \sigma_A$. $\sigma_D$ and $\sigma_A$ are the sets of diagonal and antidiagonal Pauli matrices, $\sigma_D = \lbrace \mathbb{I}, \sigma_z \rbrace$ and $\sigma_A = \lbrace \sigma_x, \sigma_y \rbrace$. Since Pauli matrices either commute or anticommute, $PQP = \left(-1\right)^{\mathcal{P}\left(\vert K \vert\right)} Q$, where $K$ is the set of pairwise anticommuting sites between $P$ and $Q$,  $\vert \cdot \vert$ is the cardinality and $\mathcal{P}$ the parity function $\mathcal{P}\left(N\right) = N \pmod 2$.

We want to show that $P$ and $Q$ commute whenever $\lbrace p_i, q_i \rbrace \in \lbrace \sigma_A, \sigma_D \rbrace, \forall 0 \leq i \leq m - 1$, i.e., if for each site $i$ the Pauli matrices in $P$ and $Q$ are from the same set, $\sigma_A$ or $\sigma_D$. Due to this restriction, the only sites that will contribute with anticommuting pairs are those where $\lbrace p_i, q_i \rbrace = \sigma_A$. Let $I = \lbrace j \vert p_j \in \sigma_A \rbrace$ be the set of sites with antidiagonal Pauli matrices. For a given Pauli word $P$ from a set with $r$ antidiagonal matrices, introduce the length-$r$ bitstring $s(P)$ that
indicates whether $\sigma_y$ was chosen at the antidiagonal sites:

\begin{equation}
    s\left(P\right)_i = \begin{cases}
        0&\text{if $p_{I_i} = \sigma_x$}\\
        1&\text{if $p_{I_i} = \sigma_y$}
    \end{cases}
    \;\forall\, 0 \leq i \leq r-1.
\end{equation}

The antisymmetry of $P$ implies an odd number of $\sigma_y$'s, so $\bigoplus_{i=0}^{r-1} s(P)_i = 1$. The parity of the number of anticommuting sites between $P$ and $Q$ is

\begin{equation}
    \bigoplus_{i=0}^{r-1}\left[s(P)_i \oplus s(Q)_i \right] = \bigoplus_{i=0}^{r-1} s(P)_i \oplus \bigoplus_{i=0}^{r-1} s(Q)_i = 1 \oplus 1 = 0.
\end{equation}
Therefore, $PQP = Q$ and $P$ and $Q$ commute.

\section{\label{sec:Sdiagonalization}Diagonalization of completely antidiagonal sets of operators}

Our goal is to find the operator that diagonalizes the set $\bigotimes_{j=1}^n \sigma_A$, with $\sigma_A = \lbrace \sigma_x, \sigma_y \rbrace$. More precisely, we want the operator that transforms $\ket{l}\bra{k} - \ket{k}\bra{l}$ into $c\left(\ket{l}\bra{l} - \ket{k}\bra{k}\right)$ with $c$ some known factor. Notice that the transformation $\sigma_y \to -i\sigma_z$ and $\sigma_x \to \mathbb{I}$ is exactly what we are looking for, with $c=1$. This is immediate to see with one site, but is also true with more than one site and any combination of diagonal and antidiagonal sites because the tensor product guarantees that only one element per row and column is non-zero, and its value is the same after the transformation.

We will now prove that the operator

\[
\Qcircuit @C=1em @R=0.7em {
& \ctrl{3} & \ctrl{1} & \gate{X} & \gate{S} & \gate{H} & \ctrl{1} & \ctrl{3}  & \qw \\
& \qw & \targ & \qw & \qw & \qw  & \targ & \qw & \qw  \\
&&&& \vdots &&&& \\
& \targ & \qw & \qw & \qw & \qw & \qw & \targ & \qw
}
\]
applies the transformation $\sigma_y \to -i\sigma_z$ and $\sigma_x \to \mathbb{I}$ but for a factor of $i$. Let us call the whole operator $V$ and decompose it in three steps, the first set of CNOTs, which we denote by $CX^{\otimes n-1}$, the local transformation $HSX$ and the final set of CNOTs. We introduce the notation $\sigma_D^0 = \mathbb{I}$, $\sigma_D^1 = \sigma_z$, $\sigma_A^0 = \sigma_x$ and $\sigma_A^1 = \sigma_y$. Given a pair of diagonal or antidiagonal operators

\begin{align}
CX\left(\sigma_D^a\otimes\sigma_D^b\right)CX &= \sigma_D^{a\oplus b}\otimes\sigma_D^b, \label{eq:cx_diagonal}\\
CX\left(\sigma_A^a\otimes\sigma_A^b\right)CX &= \left(-1\right)^{a \wedge b}\sigma_A^{a\oplus b}\otimes\sigma_D^b. \label{eq:cx_antidiagonal}
\end{align}
Here, $\oplus$ is the XOR operator and $\wedge$ the AND operator on bits.

Let $\bigotimes_{j=1}^n \sigma^{k_j}_A$ be an element of the set $\bigotimes_{j=1}^n \sigma_A$, with $k_j \in \lbrace 0,1 \rbrace$ labeling the specific antidiagonal Pauli matrix at site $j$. Applying the previous identity, the first set of transformations in $V$ is
\begin{align*}
    CX^{\otimes n-1} \bigotimes_{j=1}^n \sigma^{k_j}_A CX^{\otimes n-1} &= \left(-1\right)^{\bigoplus_{j=1}^n\bigoplus_{l=j+1}^n k_j \wedge k_l} \sigma_A^{\bigoplus_{j=1}^n k_j}\otimes\bigotimes_{j=2}^n\sigma_D^{k_j} \\
    &= \left(-1\right)^{\bigoplus_{j=1}^n\bigoplus_{l=j+1}^n k_j \wedge k_l} \sigma_A^1\otimes\bigotimes_{j=2}^n\sigma_D^{k_j},
\end{align*}
where in the second equality we have used the antisymmetry condition. This transformation is effectively changing $\sigma_y \to \sigma_z$ and $\sigma_x \to \mathbb{I}$ at all sites but the first one, which is guaranteed to be a $\sigma_y$. There is also an extra global $-1$ factor if the total number of $\sigma_y$ pairs in the original Pauli word is odd.

The next step in the transformation is $\left(HSX\right)Y\left(HSX\right)^\dagger = Z$. After the last CNOTs application, applying \ref{eq:cx_diagonal},
\begin{equation}
    V\bigotimes_{j=1}^n \sigma^{k_j}_A V^{\dagger} = \left(-1\right)^{\bigoplus_{j=1}^n\bigoplus_{l=j+1}^n k_j \wedge k_l} \sigma_D^{1\oplus\bigoplus_{j=2}^n k_j}\otimes\bigotimes_{j=2}^n\sigma_D^{k_j}
\end{equation}
Now, the first term in the tensor product $\sigma_D^{1\oplus\bigoplus_{j=2}^n k_j}$ will be $\mathbb{I}$ if $\bigotimes_{j=2}^n \sigma^{k_j}_A$ (i.e. the original Pauli word without the first site) was already antisymmetric, or $\sigma_z$ otherwise. In plain words, $V$ changes $\sigma_y \to \sigma_z$ and $\sigma_x \to \mathbb{I}$ at all sites (except for a possible $-1$ factor depending on the number of $\sigma_y$ pairs). However, this is not the goal transformation, as we wanted $\sigma_y \to -i\sigma_z$ and $\sigma_x \to \mathbb{I}$. Therefore, the transformation induced by $V$ picks an extra $i$ with respect to our goal. The antisymmetry ensures that the global factor is always $\pm i$. The sign depends on the number of $\sigma_y$ in the original Pauli word. For $3, 7, 11, \ldots$ $\sigma_y$'s, we have picked the $i$ plus an extra $i^2 = -1$ factor. for $1, 5, 9, \ldots$ $\sigma_y$'s, the possible $-1$ cancel each other. The $\left(-1\right)^{\bigoplus_{j=1}^n\bigoplus_{l=j+1}^n k_j \wedge k_l}$ factor, which is counting the number of $\sigma_y$ pairs in the original Pauli word, is $+1$ when there were $1, 5, 9, \ldots$ $\sigma_y$'s, and $-1$ in the other cases. Therefore, the factor cancels the extra $-1$'s we are picking, and $V$ is exactly the $\sigma_y \to -i\sigma_z$ and $\sigma_x \to \mathbb{I}$ transformation, but for a global factor of $i$.
\end{document}